\newcommand{\beq}{\begin{equation}}
\newcommand{\eeq}{\end{equation}}
\newcommand{\bed}{\begin{displaymath}}
\newcommand{\eed}{\end{displaymath}}
\newcommand{\beqa}{\begin{eqnarray}}
\newcommand{\eeqa}{\end{eqnarray}}
\newcommand{\beqan}{\begin{eqnarray*}}
\newcommand{\eeqan}{\end{eqnarray*}}
\newcommand{\p}{\partial}
\newcommand{\ex}{{\rm exp}}

\newcommand{\goto}{\rightarrow}

\newcommand{\ovl}{\overline}
\newcommand{\fn}{\footnote}
\newcommand{\lae}{\mathrel{<\kern-1.0em\lower0.9ex\hbox{$\sim$}}}
\newcommand{\gae}{\mathrel{>\kern-1.0em\lower0.9ex\hbox{$\sim$}}}
\newcommand{\noi}{\noindent}
\newcommand{\btab}{\begin{tabbing}}
\newcommand{\etab}{\end{tabbing}}
\newcommand{\sun}{\hbox{$\odot$}}
\newcommand{\non}{\nonumber}
\newcommand{\bfig}{\begin{figure}}
\newcommand{\efig}{\end{figure}}
\newcommand{\mx}{{\rm max}}
\newcommand{\mn}{{\rm min}}
\newcommand{\obs}{{\rm obs}}
\newcommand{\pk}{{\rm peak}}
\newcommand{\D}{{\rm D}}
\newcommand{\td}{\tilde}

\documentstyle[emulateapj,apjfonts,psfig]{article}

\slugcomment{{\sc To appear in ApJ: Oct 1, 2000}}

\begin{document}
\title{EMBEDDED, SELF-GRAVITATING EQUILIBRIA IN\\ 
SHEETLIKE AND FILAMENTARY MOLECULAR CLOUDS}
\author{CHARLES L.\ CURRY\altaffilmark{1}}
\affil{Department of Physics, 
University of Waterloo, Waterloo, ON  N2L 3G1}
\affil{and} 
\affil{Department of Physics and Astronomy, 
University of Western Ontario, London, ON  N6A 3K7}
\altaffiltext{1}{Email: curry@astro.uwo.ca} 

\begin{abstract} 
Numerical solutions of the isothermal Lane-Emden equation are 
presented, corresponding to self-gravitating gaseous cores embedded 
within a finite density envelope of overall cylindrical symmetry.  
These structures may be members of a fragmentation hierarchy proceeding 
from sheets, to filaments, to elongated, prolate clumps.  The embedded 
solutions are the first of their kind, and as such represent a significant 
improvement upon the isolated cloud paradigm used almost exclusively 
by previous authors.  The properties of the equilibria are in reasonable 
agreement with observations of dense molecular cores in star-forming 
clouds, despite the fact that there is only one free parameter in the 
models.  We show that this parameter may be identified with the 
critical wavelength for instability in the parent filament.  The 
implications of further fragmentation and the possible influence of 
magnetic fields are briefly discussed. 
\end{abstract}
\keywords{ISM: clouds --- ISM: structure --- hydrodynamics --- 
instabilities --- Stars: formation}

\section{INTRODUCTION}
\label{sec-intro}
The present observational picture of star-forming regions conveys 
cloud structures considerably more varied than any one theoretical 
scenario can explain.  Nevertheless, there are certain points of 
correspondence, when one restricts attention to the molecular, 
self-gravitating component of the interstellar medium.  
On scales of tens to hundreds of parsecs, both atomic and molecular 
gas clouds appear shell-like and filamentary (Scalo 1985; Kulkarni \& 
Heiles 1988).  Shells may be formed by several distinct dynamical 
processes: cloud-cloud collisions (Smith 1980), compressional shock 
waves from supernovae or OB stars (McCray \& Kafatos 1987), and 
large-scale shocks 
associated with spiral density waves (Roberts 1969) are but a few 
possibilities.  
Filamentary structure persists down to scales of several parsecs, 
to the regime of individual molecular clouds 
[Loren 1989, Nozawa et al 1991 ($\rho$ Oph); Heyer et al 1987, Onishi 
et al 1996 (Taurus); Bally et al 1987, Tatematsu et al 1993 (Orion)].  
Each filament typically contains several distinct subcondensations in 
close proximity to each other, some of which harbor infrared continuum 
sources (Onishi et al 1998).  In some cases, the embedded clumps are 
spaced quite regularly along the filament (Schneider \& Elmegreen 1979; 
Dutrey et al 1991).  Thus, several authors have speculated that 
the formation of star-forming clumps proceeds via a hierarchical 
fragmentation process, in which filaments are formed out 
of larger structures, and then clumps out of the filaments (Schneider 
\& Elmegreen 1979; Gaida, Ungerechts, \& Winnewisser 1984; Hanawa et 
al 1993; Fiege \& Pudritz 2000$a$). 

Focusing now on the cores themselves, maps in dense tracers such as 
NH$_3$ and CS display roughly elliptical intensity contours, with a 
mean apparent major-to-minor axis ratio of around 2 (Jijina, Myers, 
\& Adams 1999).  Statistical 
arguments applied to the distribution of measured axial ratios for 
several surveys have prompted some authors to conclude that the cores 
are more likely to be intrinsically prolate than oblate (David \& 
Verschueren 1987; Myers et al 1991; Ryden 1996).  Other statistical 
arguments imply that the observed elongation is unlikely to be a result 
of star formation or outflows (Myers et al 1991).  Coupled with the 
additional result that the majority of cores are near 
virial equilibrium (Jijina et al 1999), this argues against a wholly 
dynamical origin for prolateness, since the implied lifetimes are then  
so short that observations of cores without embedded stars would be 
exceedingly rare (the latter represent about one-half of the ammonia 
cores detected in the Taurus region, and a somewhat smaller fraction 
in Ophiuchus and Orion).  Rather, it appears more likely that cores 
are at least quasi-equilibrium structures, and that their shapes 
therefore offer some clue to the forces responsible for their 
formation. 

If this reasoning is correct, then it leads to a formidable crisis in 
our current theoretical picture of cloud equilibria, which typically 
envisions star forming cores as self-gravitating clumps bounded by a 
zero density, constant pressure---and so high temperature---medium 
(e.g., McKee et al 1993).  That is, it is difficult 
to conceive of prolate, quasi-equilibrium, gaseous cores as purely {\it 
isolated} structures.  Non-gravitational 
forces, such as rotation or magnetic fields, are not likely to aid in 
maintaining prolate equilibria, although their role in oblate structures 
has been made abundantly clear.\fn{It is in fact possible to construct 
isolated, prolate magnetic clouds, but these possess magnetic field 
structures which are highly unusual and, in any case, lack direct 
observational justification.  See Fiege \& Pudritz (2000$b$) and Curry \& 
Stahler (2000) for examples.}
The absence in the literature of even a single, physically acceptable, 
gaseous, prolate equilibrium solution suggests that a fresh theoretical 
approach is necessary.  
We examine a scenario in this paper whereby core morphology is directly 
attributable to a fragmentation process.  As we show in \S \ref{sec-plane}  
and \ref{sec-fil}, the existence of prolate structures within cylindrical 
filaments may be 
understood in exactly the same way as the existence of the filaments 
themselves: as a result of the fragmentation of the parent cloud  
and nonlinear growth of the fragments.  What has apparently escaped 
attention until now, and what we demonstrate explicitly, is that 
{\it accessible, long-lived states exist wherein cores and 
their extensive filamentary envelopes occupy the same hydrostatic 
structure}.  
We now proceed to outline a particular fragmentation hierarchy that 
might be responsible for such structures. 

\section{NONLINEAR FRAGMENTATION OF AN ISOTHERMAL LAYER}
\label{sec-plane}
To motivate the new solutions presented in the following section, 
we begin by outlining a possible formation mechanism for filamentary 
structures in the molecular interstellar medium.  
By drawing attention to selected existing results in the  
literature, we hope to provide a more compelling argument for a 
particular fragmentation hierarchy alluded to previously by several 
authors, notably Schneider \& Elmegreen (1979) and Larson (1985). 

\subsection{Analytic Solutions}
\label{sec-anal}
While not realistic in detail, a large, shell-like structure in the 
interstellar medium may be approximated on sufficiently small scales 
by a planar, self-gravitating layer, or ``sheet'' (Elmegreen \& 
Elmegreen 1978).  We restrict consideration throughout to an isothermal,
self-gravitating gas, whose equilibrium is governed by the Lane-Emden 
equation, 
\beq
\nabla^2 \psi = \frac{\p^2 \psi}{\p x^2} + \frac{\p^2 \psi}{\p y^2} +
\frac{\p^2 \psi}{\p z^2} = 4\pi G \rho_{0,s}~ e^{-\psi/a^2}, 
\label{eq-lecart}
\eeq
where $\psi$ is the gravitational potential, $a$ is the constant 
sound speed, 
and $\rho_{0,s}$ is a constant.  The equation has been written 
in cartesian coordinates $(x,y,z)$.  When $\psi$ varies solely in the 
$y$-direction, i.e.\ $\psi = \psi_{1\D} (y)$, and under the usual boundary 
conditions $\psi_{1\D} (y=0) = 0,\; (d\psi_{1\D}/dy)_{y=0} = 0$, an exact 
solution is known (Spitzer 1942):
\beq
\rho_{1\D} (y) = \rho_{0,s}~\ex [-\psi_{1\D} (y)/a^2] = \rho_{0,s}~ 
{\rm sech}^2 (y/\ell_0),  
\label{eq-spitz}
\eeq
where the scale height $\ell_0$ is defined by $\ell_0 
\equiv a/(2\pi G\rho_{0,s})^{1/2}$.
The layer is unbounded in the $y$-direction; solutions truncated by 
an external pressure at constant $y$ may also be constructed (e.g., 
Elmegreen \& Elmegreen 1978).  However, as these introduce an additional 
parameter---namely, the external pressure---into the problem, we do 
not consider them here.  The stability of the unbounded solution 
was examined by Ledoux (1951), who found gravitational instability for 
infinitesimal 
sinusoidal perturbations of wavelength exceeding $\lambda_{x,{\rm cr}} 
= 2\pi \ell_0$.  The maximum growth rate occurs at $\lambda_{x,{\rm MGR}} = 
2.24 \lambda_{x,{\rm cr}} = 14.1 \ell_0$.  The choice of the $x$-direction for 
the perturbation wavevector here is arbitrary; the same results hold for  
perturbations in the $z$-direction only. 

Subsequently, Schmid-Burgk (1967) (hereafter S-B) presented a remarkable 
{\it two-dimensional} (2D) solution of equation (\ref{eq-lecart}) 
(independently discovered, in a hydrodynamic context, by Stuart 1967):
\beq
\rho_{2\D} (x,y) = \rho_{0,s} ~\ex \left[\frac{-\psi_{2\D} (x,y)}{a^2}
\right] = \frac{\rho_{0,s} 
~(1 - A^2)}{[{\rm cosh}~ (y/\ell_0) - A~ {\rm cos}~ (x/\ell_0)]^2}. 
\label{eq-sbsn}
\eeq
The constant $A,~ 0 < A < 1$ describes the amplitude of spatially 
periodic variations along one direction parallel to the layer 
(here chosen as $x$), with the same critical wavelength, 
$\lambda_{x,{\rm cr}}$, 
as found by Ledoux.  As noted by S-B, this result shows that the 
latter mode is not restricted to infinitesimal amplitudes.  Indeed, 
solutions of {\it all} amplitudes $A$ are contained in equation 
(\ref{eq-sbsn}).\fn{Note also that, like other solutions of the 
Lane-Emden equation,  S-B's solution is homology invariant 
(Chandrasekhar 1958); i.e.\ the potential $\psi^*_{2\D} \equiv 
\psi_{2\D} (Cx,Cy) 
- 2 ~{\rm log}~ C$, $C =$ constant, is also a solution of equation 
(\ref{eq-lecart}).}  It is noteworthy that the periodicity of this exact, 
nonlinear solution does not correspond to that of the fastest-growing 
mode in the linearly unstable layer, but rather to the {\it critically} 
unstable mode.  This feature is discussed further in \S \ref{sec-disc}. 

As $A \goto 0$, equation (\ref{eq-sbsn}) reduces to Spitzer's 
solution.  In the limit $A \goto 1$, the equilibrium\fn{The term 
``equilibrium'' is used here in its strictly limited sense: i.e., 
a particular equilibrium may be either stable or unstable to 
perturbations in a direction in which the solution is 
translation-invariant (in this case, the $z$-direction).}
corresponds 
to a series of parallel cylindrical ``fragments,'' infinitely 
extended in the $z$-direction, with density maxima separated by a 
distance $\lambda_{x,{\rm cr}}$.  For general $A$, the fragments 
are embedded, nearly elliptic cylinders with equidensity contours 
of eccentricity $(1 - A)^{1/2}$ (see Figure 1 of S-B).  The density 
maxima and minima along the $x$-axis (each spaced at an interval of 
$\Delta x = 2\pi \ell_0$), are given by 
\beq
\rho_c = \rho_{0,s} \left(\frac{1 + A}{1 - A}\right) ~~~~~ 
{\rm and} ~~~~~ \rho_s = \rho_{0,s} \left(\frac{1 - A}{1 + A}
\right),
\label{eq-maxmin}
\eeq
respectively. 

\subsection{Numerical Solution}
\label{sec-numer1}
There is reason to suspect that analogues of the S-B solution 
exist in other geometries, in particular in cylindrical symmetry, 
a case we shall focus on in the following section.  Unfortunately, 
to our knowledge, no other embedded solutions are known in analytic 
form.  However, numerical techniques can be used.  As a test of one 
such scheme used later in this paper, we attempted to find the S-B 
solution using the self-consistent field method (e.g.\ Tassoul 1978).  
That is, equation (\ref{eq-lecart}) was solved iteratively 
on a two-dimensional grid, subject to the boundary conditions
\beqa
\left .\frac{\p \psi'}{\p x'}\right|_{x' = 0} &=& 
\left .\frac{\p \psi'}{\p y'}\right|_{y' = 0} = 0, 
\label{eq-bc1} \\
\psi' (0,0) = -{\rm ln}~\rho' (0,0), ~~~~~~~~~~ 
&{\rm and}& ~~~~~~~~~~ 
\left .\frac{\p \psi'}{\p x'}\right|_{x' = X'} = 0. 
\label{eq-bc2}
\eeqa
Here a prime indicates a dimensionless quantity---e.g., $\psi' \equiv 
\psi/a^2$---and $0 \leq x' \leq X',~ 0 \leq y' \leq Y'$ is the extent 
of the computational region.  The nondimensionalization 
of length will be discussed below; for now it may be assumed arbitrary. 

The first 
of conditions (\ref{eq-bc1}) ensures that the $x$-component of the 
gravitational field vanishes on the $y$-axis, while the second imposes 
reflection symmetry about the $x = 0$ plane.  Conditions (\ref{eq-bc2}) 
are consistent with the Spitzer solution, although neither constrains 
the numerical solution to be identical to the former in any particular 
limit.  In addition, these conditions allow for solutions with a periodic 
structure in $x$.  In an attempt to find these, we took $\psi'_{1\D,0} 
(y') = -{\rm ln}~\rho'_{1\D,0} (y')$ (see eq.\ \ref{eq-spitz}) as an 
initial guess, and added a perturbation of the form
\bed
\delta \psi' (x', y') = \epsilon~ {\rm cos} (p x'/X')~ {\rm sin} 
(p y'/Y'), 
\eed
where $\epsilon$ is a small, constant amplitude and $0 < p < 2\pi$.  
For convergence of the numerical code we required 
$1 - \psi'^{(n)}/\psi'^{(n+1)} < \delta = 0.005$, where $\psi'^{(n)}$ 
is the value of $\psi'$ at the $n^{\rm th}$ iteration.  
For $\epsilon \lae 0.005$, the method converged immediately to the 
Spitzer solution (\ref{eq-spitz}).  For any larger 
$\epsilon$, however, a unique, 2D structure resulted.  The equilibria 
so obtained constitute a family of solutions in the single parameter 
$X'$, such that the density contrast $\rho_c/\rho_s$ increases 
monotonically with increasing $X'$.  Here $\rho_s \equiv \rho(X,0)$.  
The vertical extent $Y'$ was chosen sufficiently large that the effect 
of different $Y'$ on 2D solutions of the same $X'$ was negligible.  

In order to compare these solutions to the analytic one of S-B, we 
need to specify the nondimensionalization.  This is, in fact, already 
implicit in the solution method.  At each iteration, the Lane-Emden 
equation is solved for $\psi' (x',y')$ with the source term 
\bed
\rho'^{(n)} (x',y') = \ex [\psi'^{(n)} (0,0) - \psi'^{(n)} (x',y')]. 
\eed
Thus, $\rho' (x',y')$ is normalized with respect to its {\it central} 
value at each iteration; i.e.\ $\rho'^{(n)} (0,0) = 1$ (this is made 
explicit in the first of equations \ref{eq-bc2}).  Consequently, 
once an exact, 2D equilibrium is found, all ``memory'' of the original 
density scale $\rho_{0,s}$ is lost.  The characteristic 
density of the solution is $\rho_c = \rho (0,0)$, and thus the 
corresponding unit length is $\ell_c \equiv a/(2\pi G \rho_c)^{1/2}$,   
not $\ell_0 \propto \rho_{0,s}^{-1/2}$ as in the analytic solution. 
Hence, the appropriate nondimensionalization is specified a posteriori 
as
\beq
(x',y',z') \equiv (x/\ell_c, y/\ell_c, z/\ell_c). 
\label{eq-nondim1}
\eeq

An exact relation between the parameter $X'$ in the numerical 
problem, and the amplitude $A$ in the analytic solution, may now be 
derived.\fn{I thank Dean McLaughlin for pointing out a simplified 
derivation.}  Solving equation (\ref{eq-sbsn}) for $\psi_{2\D} (x,y)$ 
and substituting the result into the second of equations (\ref{eq-bc2}) 
gives
\beq
X' = \frac{X}{\ell_c} = \pi \left(\frac{\rho_c}
{\rho_s}\right)^{1/4} = \pi \left(\frac{1+A}{1-A}\right)^{1/2}. 
\label{eq-xamp}
\eeq
The parameter $X'$ represents one-half of the spacing between the 
embedded fragments; i.e., $\lambda'_x = 2X'$.
In the analytic solution, the spacing of the fragments is constant, 
i.e.\ $\Delta x/\ell_0 = \lambda_{x,{\rm cr}}/\ell_0 = 2\pi$.  There 
the fragment spacing stays the same while the amplitude $A$ varies.  
Just as analytic solutions exist for all $A$, numerical solutions 
exist for all values of $X' > \pi$.

Figure 1 of S-B (1967) shows a solution with $A = 0.17$, whose  
corresponding density contrast is $\rho_c/\rho_s = 1.987$.  Inserting 
this $A$ into equation (\ref{eq-xamp}), one finds $X' = 3.730$. 
Using this $X'$ as input for the numerical method then yields a 
solution with $\rho_c/\rho_s = 1.981$, a difference of 0.3 percent 
from S-B's analytic result.  This solution is displayed in Figure 1.
The discrepancy between the analytic and numerical solutions remains 
small as $X'$ increases above 3.73, and is still less than 1 
percent for $X' = 22,~ \rho_c/\rho_s \approx 2400$.  At smaller $X'$, 
the numerical and analytic results slowly diverge, with the relative 
error in $\rho_c/\rho_s$ reaching 5 percent at $X' = 3.32$, and  
nearly 20 percent at $X' = \pi$.  At the latter value, equation 
(\ref{eq-xamp}) predicts $A = 0$ and $\rho_c/\rho_s = 1$.  Presumably, 
the method is not sufficiently sensitive to the slight density contrasts 
present in these small amplitude ($A \lae 0.05$) solutions to render an 
accurate result.  On the other 
hand, for $X' < 3.125$, the numerical method converges to the 
Spitzer solution, with a slightly non-uniform density along the midplane 
(e.g., $\rho_c/\rho_s \simeq 0.99$ at $X' = 3.12$).  
Overall, these results confirm the validity of the numerical method, 
with the nondimensional length and density chosen as in equation 
(\ref{eq-nondim1}).  We may expect the same to be true for other 
solutions found by the same technique, irrespective of whether an 
analytic solution is known. 

We conclude our discussion of the S-B solution by noting that 
related states have in fact appeared in the literature, in the 
context of either linear perturbation theory (Miyama, Narita, \& 
Hayashi 1987$a$) or time-dependent nonlinear calculations 
(Miyama, Narita, \& Hayashi 1987$b$).  The basic character of the 
solution persists even in the presence of a ($z$-independent) 
magnetic field (Fleischer 1998; Nagai, Inutsuka, \& Miyama 1998), 
or when the layer is truncated by an external medium.  The 
pressure-bounded, magnetized layer has additional modes of 
fragmentation available to it, as shown by Nagai et al (1998).   
However, the significance of the basic result---namely, that the 
embedded filaments are in fact {\it exact, equilibrium 
solutions}---is rarely emphasized.

\section{NONLINEAR FRAGMENTATION OF AN ISOTHERMAL FILAMENT}
\label{sec-fil}
\subsection{Equilibrium and Stability}
\label{sec-equi}
In cylindrical coordinates $(r,\phi,z)$, the isothermal Lane-Emden 
equation reads
\beq
\nabla^2 \psi = \frac{1}{r}\frac{\p}{\p r}\left(r \frac{\p \psi}{\p r}
\right) + \frac{\p^2 \psi}{\p z^2} = 4\pi G \rho_{0,f}~e^{-\psi/a^2},
\label{eq-lecyl}
\eeq
where $\rho_{0,f}$ is a constant. 
Here the $z$-axis is taken to coincide with the principal axis of the 
cylinder, and we have taken $\p/\p \phi = 0$, restricting consideration 
to axisymmetric solutions.  When $\psi$ depends only on $r$, i.e.\ 
$\psi = \psi_{1\D} (r)$, and under the 
boundary conditions $\psi_{1\D} (r=0) = 0,\; (d\psi_{1\D}/dr)_{r=0} = 0$, 
Stodolkiewicz (1963) and Ostriker (1964) derived the following exact 
solution of equation (\ref{eq-lecyl}) (hereafter the S-O solution): 
\beq
\rho_{1\D} (r) = \rho_{0,f}~\ex \left[\frac{-\psi_{1\D} (r)}{a^2}
\right] = \frac{\rho_{0,f}}{(1 + r^2/l_0^2)^2},
\label{eq-sosoln}
\eeq
where $l_0 \equiv (2a^2/\pi G\rho_{0,f})^{1/2}$ is the cylindrical 
scale radius.  Interestingly, solution (\ref{eq-sosoln}) reduces to 
that of S-B as $A \goto 1$ in the latter; details may be found in the 
Appendix.  This suggests that equilibria quite similar to the S-O 
solution may result directly from the fragmentation of an isothermal 
layer. 

The solution (\ref{eq-sosoln}) decreases as $r^{-4}$ at large $r$; 
as in the planar case, solutions truncated by an external pressure at 
constant $r$ have also been considered (e.g., Inutsuka \& Miyama 1997; 
Fiege \& Pudritz 2000$a$).  The maximum mass per unit 
length of the filament described by equation (\ref{eq-sosoln}) 
is\fn{In the S-B solution, the total mass per unit $z$-length of each 
``cell'', $-\pi \leq x/l_0 \leq \pi$, is also $2a^2/G$.} 
\beq
\mu_\mx \equiv 
\int_0^{\infty} 2\pi \rho_{1\D} (r) r dr = \frac{2a^2}{G}.
\label{eq-linem}
\eeq
Isothermal cylinders with density distributions having $\mu > \mu_\mx$ 
are unstable to radial collapse, while those with $\mu < \mu_\mx$ 
expand radially outward, unless confined by an external pressure 
(e.g., Inutsuka \& Miyama 1992). 
The stability of solution (\ref{eq-sosoln}) to axisymmetric, 
linear perturbations was examined by Stodolkiewicz (1963), who found 
instability for perturbations of wavelength exceeding 
$\lambda_{z,{\rm cr}} = 
3.94 ~l_0$.  The maximum growth rate occurs at $\lambda_{z, {\rm MGR}} 
= 1.98 ~\lambda_{z,{\rm cr}} = 7.82 ~l_0$ (Nagasawa 1987).  Further 
analysis has been carried out by Inutsuka \& Miyama (1997) in the 
nonlinear regime of perturbation growth.  However, the nonlinear 
{\it resolution} of the instability 
remains an open question.  That is, what is the final outcome of the 
fragmentation instability in an isothermal cylinder? 

\subsection{Numerical Solution}
\label{sec-numer2}

We undertake a numerical investigation of the question posed above, 
using an identical technique to that described in \S \ref{sec-numer1}. 
That is, rather than following the time evolution of a particular 
unstable state, we instead search for an exact, static solution having 
a 2D structure.  The boundary conditions on the potential are now
\beqa
\left .\frac{\p \psi'}{\p r'}\right|_{r' = 0} &=& 
\left .\frac{\p \psi'}{\p z'}\right|_{z' = 0} = 0, 
\label{eq-bcp1} \\
\psi' (0,0) = -{\rm ln}~\rho' (0,0), ~~~~~~~~~~ 
&{\rm and}& ~~~~~~~~~~ 
\left .\frac{\p \psi'}{\p z'}\right|_{z' = Z'} = 0,
\label{eq-bcp2}
\eeqa
where $0 \leq r' \leq R',~ 
0 \leq z' \leq Z'$ is the size of the computational region.  
The nondimensionalization is the same as that used in \S 
\ref{sec-numer1}, except with $\ell_c$ replaced by 
$l_c \equiv (2a^2/\pi G\rho_c)^{1/2}$, where $\rho_c \equiv 
\rho (0,0)$.  Again we took the equilibrium potential, 
$\psi'_{1\D} (r') = -{\rm ln}~\rho'_{1\D} (r')$
as an initial guess, and added a perturbation of the form 
\bed
\delta \psi' (r',z') = \epsilon~ {\rm sin} (p r'/R')~ {\rm cos} 
(p z'/Z'),
\eed
with $0 < p < 2\pi$.  
For any $\epsilon \gae 0.005$, a family of unique 2D structures was 
again found, now parameterized by $Z'$. 
The corresponding $R'$ was chosen sufficiently large that its effect 
on the solutions was negligible.  In practice, choosing $R'$ to be 
approximately twice the radius of the ``tidal lobe''---i.e., the last 
closed isodensity contour, of density $\rho_s \equiv \rho(0,Z')$---was 
sufficient. 

The density contrast between the center and the tidal lobe, $\rho_c/
\rho_s$, is a monotonically increasing function 
of $Z'$.  Models with $1.2 < \rho_c/\rho_s \lae 10^3$ were generated, 
corresponding to $2.1 < Z' \leq 30$.  A few representative equilibria are 
displayed in Figure \ref{fig-plegs}.  At small $Z'$, the equidensity 
contours are highly prolate (Fig.\ \ref{fig-plegs}$a$); as $Z'$ increases, 
the interior contours become nearly spherical (Fig.\ \ref{fig-plegs}$d$). 
Note that we were able to find 2D equilibria only down to a minimum $Z' 
\simeq 2.1$; below this value, 
convergence was not obtained until $Z'=1.8$.  For all $Z' \leq 1.8$, 
the method converged to the S-O 
solution, with a slightly non-uniform density along the axis of symmetry 
(e.g., $\rho_c/\rho_s = 0.997$ at $Z' = 1.8$).  As we discuss below in 
\S \ref{sec-disc}, these results strongly suggest a fragmentation scale 
$\simeq \lambda_{\rm cr}$, as in the planar layer. 

Physically, the quantity $2Z'$ may be interpreted as the separation between 
any two clumps in a linear chain of identical condensations (e.g.\ Lizano 
\& Shu 1989; Fiedler \& Mouschovias 1992).  This feature thus mimics 
an observed property of star-forming environments: namely, that pre-stellar 
cores are rarely found in isolation.  In \S \ref{sec-plseq}, we examine how 
the properties of the condensations depend upon $Z'$, and how the latter 
may be constrained by observations.  

\section{PHYSICAL PROPERTIES OF THE FRAGMENTS}
\label{sec-pl2}

\subsection{Virial Theorem Analysis}
\label{sec-virial}
Insight into the global properties of equilibria may be obtained 
from the scalar virial theorem.  In the present context of embedded 
cores, however, care must be taken regarding the region of 
application.  For an axisymmetric cloud in equilibrium, we have
(McKee et al 1993) 
\beq
2U + \Pi + W = 0,
\label{eq-virial}
\eeq
where
\beqan
U & = & \frac{3}{2}\int_V P dV = \frac{3}{2} a^2 M, \\
\Pi & = & -\int_S P {\bf r}\cdot {\bf dS}, 
\eeqan
and
\bed
W = -\int_V \rho ~{\bf r}\cdot {\bf \nabla}\psi dV
\eed
are the relevant thermal, compressive, and gravitational energies. 
These integral quantities are usually summed over the volume $V$ 
enclosed within an arbitrary closed surface $S$.  In the present 
case, one might choose for $S$ any closed isobar.  Then the thermal 
and compressive terms are easily calculated.  However, for any choice 
of $S$ the gravitational energy $W$ is incomplete because part of the 
gravitational field arises from {\it outside} the region considered.  
This is in keeping with the nature of embedded equilibria.  Indeed, 
each equilibrium found by the numerical method of the previous 
section comprises an entire ``cell,'' i.e., the region 
$0 \leq r \leq R,\; 0 \leq \phi \leq 2\pi,\; -Z \leq z \leq Z$, 
so that the above choice of $S$ is inappropriate in any case.  
The virial theorem may only be consistently applied on the same 
region.  Since the gravitational force vanishes on the plane circular 
section $z=\pm Z$ (by the second of the boundary conditions eq.\ 
\ref{eq-bcp2}), the corresponding energy $W$ summed over the entire 
cell between $z=-Z$ and $z=+Z$ is therefore complete. 

Figure \ref{fig-virial} shows the behavior of individual terms in the 
virial equation (\ref{eq-virial}), expressed in dimensionless units, 
along with their sum, $\Sigma \equiv 
2U + \Pi + W$, as one proceeds along a sequence of increasing central 
density $\rho_c/\rho_s$.  Here, $\rho_s$ is used as a convenient 
reference density only; the virial terms have 
been calculated for the entire cell, not just for the region inside 
the tidal lobe.  The figure shows that the compressive term $\Pi$ has 
a magnitude of only 40 to 50 percent of the gravitational term $W$ over 
most of the sequence.  This may be compared with the case of isolated 
clouds, where external compression exceeds the effect of self-gravity
at low density contrast, the latter becoming dominant only for more 
centrally-condensed clouds.  There, the boundary pressure is all that 
prevents the cloud from dispersing.  That there are no such equilibria in 
the present sequence can also be seen by considering the total energy, 
$E \equiv W + U$, also plotted in Fig.\ \ref{fig-virial}.  Since 
$E < 0$ for all $\rho_c/\rho_s$, {\it all of the equilibria are 
gravitationally dominated.}  Indeed, this feature is consistent with 
the notion that such structures are the result of nonlinear fragmentation 
within the self-gravitating parent filament.  Note that this also means 
that the prolate shape of the cores is {\it not} 
primarily due to tidal stretching by adjacent cores; an effect that,  
in the isolated case, may be mimicked by use of the boundary condition 
eq.\ (\ref{eq-bcp2}) (see, e.g., Curry \& Stahler 2000). 
Finally, note that all three of $U, ~\Pi$ and $W$ approach nonzero 
limiting values as $\rho_c/\rho_s \goto 1$.  This is to be expected, 
since these quantities are nonzero in the S-O cylinder.  Their limiting 
values are readily calculated as: $U_0 = 3\pi Z',~ \Pi_0 = -2\pi Z'$, and 
$W_0 = -4\pi Z'$.

\subsection{A Sequence of Prolate Equilibria}
\label{sec-plseq}
As indicated by the dashed curves in Fig.\ \ref{fig-plegs}, for a given 
$Z'$ there exists a unique tidal lobe, within which the equidensity 
countours are closed, and that corresponds to a density minimum along 
the symmetry axis, $r = 0$.  The intersection of the tidal lobe and the 
$z=0$ plane is a circle whose radius we denote by $R_t$.  The 
tidal lobe is thus one possible 
definition of the ``surface'' of the dense core, since it marks where 
the latter may be distinguished from the background filament.  However, 
these solutions make clear the danger of taking such a term too 
literally.  On this interpretation, 
we may fix the density on the tidal lobe at a value appropriate to the 
intercore medium.  (Defining the extent of a core using a single 
equidensity contour is in fact in accord with the usual observational 
definition, which assigns a core's size on the basis of its associated 
half-power intensity contour).  

Fixing the intercore density $\rho_s = \ovl{m} n_s$ and temperature 
$T$ allows one to characterize the size and mass of the fragments via 
the following reference quantities:
\beq
l_s \equiv \left(\frac{a^2}{G \rho_s}\right)^{1/2} = 
\left( \frac{\pi}{2} \frac{\rho_c}{\rho_s}\right)^{1/2} l_c,~~~~~~~ 
m_s \equiv \rho_s l_s^3 = \frac{a^3}{(G^3 \rho_s)^{1/2}}.  
\label{eq-lsms}
\eeq
In terms of these reference values, we now define the dimensionless 
lengths and mass:
\beq
\td{Z} \equiv \frac{Z}{l_s} = Z' \left(\frac{\pi}{2}\frac{\rho_c}
{\rho_s}\right)^{-1/2},~~~~ \td{R}_t \equiv \frac{R_t}
{l_s}, ~~~~{\rm and}~~~~~~~~ \td{M}_t \equiv 
\frac{M_t}{m_s}, 
\label{eq-nondim2}
\eeq
where $M_t$ is the dimensional mass contained within the tidal 
lobe.  

Figure \ref{fig-plseq} shows the behavior of $\td{Z},~ \td{R}_t$, 
and $\td{M}_t$ as a function of density contrast for fixed $a$ and 
$n_s$.  The results are reminiscent of the Bonnor-Ebert (hereafter B-E) 
sequence, shown by dashed lines, but with a maximum in mass at 
$\rho_c/\rho_s \simeq 10.0$ instead of at $(\rho_c/\rho_s)_{\rm BE} 
= 14.04$.  The maximum mass of the prolate sequence, $\td{M}_t 
= 1.32$, 
exceeds that of the B-E sequence, $M_{\rm BE} = 1.182$ by 12 percent. 
The polar radius $\td{Z}$ decreases from a value of 1.53 at 
$\rho_c/\rho_s \simeq 1.2$ to a minimum of 0.60 at $\rho_c/\rho_s 
\simeq 600$.  The tidal lobe radius $\td{R}_t$ has a 
maximum of 0.45 at $\rho_c/\rho_s \simeq 6.4$, and a minimum of 0.34 
at $\rho_c/\rho_s \simeq 10^3$.  

Finally, it is instructive to compare the density structure of the 
prolate fragments with that of known equilibria.  The equatorial 
and polar density profiles for the models displayed in 
Figs.\ \ref{fig-plegs}$c$ and $d$ are shown in Figures \ref{fig-plrho}$a$ 
and $b$.  These figures also show the radial density profiles of a S-O 
cylinder and of a marginally stable B-E sphere.  At the relatively 
low $\rho_c/\rho_s$ of Fig.\ \ref{fig-plrho}$a$, the polar density profile 
is shallower than $z^{-2}$, while the equatorial density resembles that of 
a B-E sphere out to about three times its ``core radius,'' after which 
it steepens to the $r^{-4}$ of the S-O profile.  At high $\rho_c/\rho_s$ 
(Fig.\ \ref{fig-plrho}$b$), the similarity of both density profiles to 
that of the B-E sphere is evident.  Thus, except at very low central 
concentrations or at distances far outside the tidal lobe, the core's 
density structure bears little resemblance to that of its parent S-O 
cylinder.  This is consistent with observations of globular filaments, 
which rarely imply density profiles steeper than $r^{-2}$ (Alves et 
al 1998; Johnstone \& Bally 1999; Lada, Alves, \& Lada 1999).

\subsection{Remarks on Stability}
\label{sec-stab}
In the case of isolated, pressure-bounded clouds, the presence 
of a maximum in cloud mass as one proceeds from smaller to larger 
$\rho_c/\rho_s$ signifies a transition from stable to unstable 
equilibria.  However, this technique (often referred to as the 
``static method''; Tassoul 1978), cannot be carried over to the 
embedded cores characterized by the mass $\td{M}_t$, since they 
are not complete equilibria (\S \ref{sec-virial}).  Thus, the 
maximum in the $\td{M}_t$ vs.\ $\rho_c/\rho_s$ relation seen in 
Fig.\ \ref{fig-plseq}, while suggestive, has no direct bearing on 
the issue of stability. 

Another method used to investigate the stability of isolated clouds 
is that of the Gibbs free energy (Stahler 1983; 
Tomisaka et al 1988).  Unfortunately, this method is also unlikely 
to succeed here, for the following reason.  Along a sequence of 
equilibria of fixed temperature and surface pressure, the Gibbs free 
energy is a minimum at the critical stability point.  However, in 
the present context, an equilibrium consists of an entire cell, which  
lacks a single, isobaric bounding surface.  It is therefore 
impossible to construct a sequence of cells 
of fixed $a$ {\it and} $n_s$, for which quantities such as mass  
have definite maxima.  Consequently, one finds that the Gibbs free 
energy is a monotonically decreasing function of $\rho_c/\rho_s$; 
i.e., no minimum value is attained along the sequence.  
Thus, the free energy approach cannot be applied in the usual manner.  
Explicit, time-dependent calculations of the evolution of small 
perturbations within the equilibria may be required to illuminate 
this important issue.

\section{COMPARISON WITH OBSERVATIONS}
\label{sec-comp}
To compare with observations, it is useful to write $l_s$ and $m_s$ as
\beqa
l_s & = & 0.382~ {\rm pc}~\left(\frac{a}{0.19~ {\rm km~ s}^{-1}}\right) 
\left(\frac{n_s}{10^3 ~{\rm cm}^{-3}}\right)^{-1/2}, 
\label{eq-ldim} \\
m_s & = & 3.19~ M_{\sun}~\left(\frac{a}{0.19~ {\rm km~ s}^{-1}}\right)^3 
\left(\frac{n_s}{10^3 ~{\rm cm}^{-3}}\right)^{-1/2}. 
\label{eq-mdim}
\eeqa
The scaling for $n_s$ is the minimum total number density, $n = 
1.2 n_{H_2}$, estimated from $^{13}$CO measurements of filamentary 
clouds (Nercessian et al 1988), while that for $a$ is the thermal 
sound speed for $T = 10$ K and a mean molecular weight of $\ovl{m} 
= 2.33~m_H$.  
In a high mass star-forming region 
such as Orion, $n_s$ can be as high as $\sim 10^4$ cm$^{-3}$ 
(Dutrey et al 1993), while $a \propto T^{1/2}$ can exceed 0.40 
km s$^{-1}$.  Nonthermal motions, which dominate thermal motions 
in clouds above scales $\sim$ few $\times ~0.1$ pc, may be included 
in a schematic manner by replacing $a$ with a constant velocity 
dispersion $\sigma$.  Representative values of $\sigma$ are 
given below.  

With the above scalings, the results of Fig.\ \ref{fig-plseq} 
correspond to the following ranges:
\beqan
0 < M/M_{\sun} \leq 4.2, \\
0.23 ~{\rm pc} \leq Z \leq 0.58 ~{\rm pc}, \\
0 \leq R_t \leq 0.17 ~{\rm pc}. 
\eeqan
These ranges are in reasonable agreement with deduced values for 
cores observed in dense tracers, such as ammonia (Jijina et al 1999). 
This is true even though NH$_3$ has a critical excitation density, 
$n_{\rm ex} \sim 10^4$ cm$^{-3}$, an order of magnitude larger than 
our assumed $n_s$.  However, this effect is offset in equations 
(\ref{eq-ldim}) and (\ref{eq-mdim}) by the fact that some regions 
have $T > 10$ K and that all exhibit nonthermal motions. 

We now compare the intercore separations found in this simple model 
with those observed in star-forming regions.  Two values 
of $\td{Z}$ may be singled out from Fig.\ \ref{fig-plseq} as being of 
particular interest.  The first, denoted by $\td{Z}_\pk$, 
corresponds to the peak in $\td{M}$, and has the value 
$\td{Z}_\pk = 0.94$.  The second notable value is the minimum 
in $\td{Z}$ occuring near the right-hand side of Fig.\ 
\ref{fig-plseq}, $\td{Z}_\mn = 0.60$.  The dimensional value of each 
of these quantities depends upon both the intercore density and 
velocity dispersion through equations (\ref{eq-nondim2}) and 
(\ref{eq-ldim}).  Figure 
\ref{fig-plic} shows both characteristic separations as a function 
of $n_s$ for various $\sigma$, where the latter is given in terms 
of $a_{10} \equiv 0.19$ km s$^{-1}$, the thermal sound speed in $T=10$ 
K gas.  The range of observed intercore separations, $2Z_\obs$, in 
each of three star-forming regions---Taurus, Ophiuchus, and Orion---
is indicated by vertical bars on the graph.  These ranges were 
deduced from molecular line maps of cores embedded within filaments 
in each region.  The characteristic intercore density  
was estimated as the observed density at a scale corresponding to the 
mean value of $2Z_\obs$ for that region.  Further details on the data 
used may be found in the caption to Table 1.  

Figure \ref{fig-plic} shows that the observed separations agree with 
$2Z_\pk$ for $\sigma/a_{10} \simeq 1-2.5$ in Ophiuchus and 
Orion, and for $\sigma/a_{10} \simeq 1.5-3.5$ in Taurus.  The 
corresponding ranges computed for $2Z_\mn$ are $\sigma/a_{10} \simeq 
1.5-4$ in Ophiuchus and Orion, and $\sigma/a_{10} \simeq 2-5$ in 
Taurus.  These ranges may now be compared with the observed intercore 
velocity dispersion $\sigma_\obs$ in each region.  
In Taurus, the range computed according to $2Z_\pk$ 
includes $\sigma_\obs$, while in Ophiuchus and Orion, $\sigma_\obs$ 
lies above the range of theoretical values.  On the other hand, the 
ranges computed for $2Z_\mn$ include $\sigma_\obs$ for all three regions.  
However, since the physical relevance of $Z_\mn$ is highly uncertain due 
to unresolved stability issues (\S \ref{sec-stab}), henceforth we focus 
on the results for $Z_\pk$.  

The discrepancy obtained using $Z_\pk$ can be stated alternatively as 
follows: given the observed $\sigma$ and $n_s$ in both Ophiuchus and 
Orion, the model appears to overestimate
$2Z_\obs$.  These estimates, $2Z_{\rm pred}$, are given in Table 1 
and plotted as open circles in Fig.\ \ref{fig-plic}.  However, this 
disagreement may have a rather simple explanation. 
The model-derived values assume that the parent filament is aligned 
exactly perpendicular to the line of sight---an unlikely circumstance. 
Generally, the filament will be inclined at some angle $90^\circ - i$ 
to the line of sight, so that $2Z_{\rm pred}$ is reduced to 
$2Z_{\rm pred}$cos$\;i$.  To bring this into agreement with the 
mean $2Z_\obs$ then requires $i = 50^\circ$ in Ophiuchus and 
$i = 53^\circ$ in Orion. 

An independent measure of the intercore velocity dispersion may 
be obtained from the observed properties of filaments.  According 
to equation (\ref{eq-linem}) with $\sigma$ in place of $a$, the 
maximum line mass of the parent filament from which the cores could 
have condensed is:
\beq
\mu_\mx = 16.4~ \left(\frac{\sigma_\mx}{a_{10}}\right)^2 
\frac{M_{\sun}}{\rm pc}.
\label{eq-linem2}
\eeq
Using available observations, we computed an approximate line mass, 
$\mu_\obs$, for filaments in Taurus, Ophiuchus, and Orion, with the 
results given in Table 1.  The corresponding $\sigma_\mx$, obtained 
from equation 
(\ref{eq-linem2}) assuming $\mu_\obs = \mu_\mx$, is also tabulated. 
In Taurus and Ophiuchus, $\sigma_\mx$ and $\sigma_\obs$ are in 
close agreement, indicating that the filaments themselves are near 
virial equilibrium.  In Orion, $\sigma_\mx$ should not be compared 
with $\sigma_\obs$, since the latter is obtained from C$^{18}$O 
measurements, whereas $\mu_\obs$ is derived from $^{13}$CO 
observations (Bally et al 1987).  Instead, we note that the 
corresponding linewidth, $\Delta V (^{13}{\rm CO}) = 2.2$ 
kms$^{-1}$, gives a dispersion of $\sigma_\obs (^{13}{\rm CO}) 
= 0.97$ kms$^{-1}$, again in very good agreement with $\sigma_\mx$.

\section{DISCUSSION}
\label{sec-disc}
If the effect of inclination alone does not account for the above 
discrepancy between $2Z_\pk$ and $2Z_{\rm pred}$, then 
this may indicate a more fundamental limitation of the model in its 
present form.  The smaller observed intercore separation 
(for a given $n_s$ and $\sigma$) might be the result of further 
fragmentation of {\it stable} cores lying at the low-density end of 
the equilibrium sequence, Fig.\ \ref{fig-plseq}.  However, the 
resulting cores would have correspondingly smaller masses than those 
obtained above in \S \ref{sec-comp}, which are already in only 
marginal agreement with observations.   

Although this paper has not included magnetic effects, it is worth 
mentioning that, in the presence of a longitudinal magnetic field $B$
that decreases in strength outward from the parent cylinder axis, the 
critical wavelength for instability is significantly reduced
(Nakamura, Hanawa, \& Nakano 1993; Hanawa et al 1993).  This occurs 
because the effective sound speed, $c_{\rm eff} = (a^2 - v_A^2)^{1/2}$, 
where $v_A \equiv B/(4\pi G \rho)^{1/2}$ is the Alfv\'en speed, is 
decreased from its non-magnetic value, thereby reducing 
thermal support against gravity in the longitudinal direction.  
If a suitable generalization of the solutions found here exists for 
magnetized, isothermal cylinders, then the constituent fragments 
should have a smaller spacing for a given density contrast, perhaps 
offering more satisfactory agreement with observations.  

What is the relation between the intercore spacing and the most 
unstable and critical wavelengths in the S-O cylinder?  We saw in \S 
\ref{sec-plane} that, in the case of the isothermal layer, it is the 
{\it critical} wavelength that determines the inter-filament spacing 
in S-B's 2D solution.  This feature merits further comment.  Let us 
imagine the temporal growth of an unstable fragment, either filamentary  
or prolate, to proceed as follows.  Fragmentation begins at the scale of 
$\lambda_{\rm MGR}$, the wavelength of the fastest-growing linearly 
unstable mode of the parental gas distribution.  As the fragment (defined 
as the material residing within the tidal lobe) grows into the nonlinear 
regime, it begins to attract, and be attracted by, surrounding fragments 
with the same properties.  Thus, the entire chain of masses becomes more 
tightly bound as a result of the increased gravitational potential energy 
due to condensation, and the intercore spacing decreases.  There is a 
limit to this process, however, at the scale of $\lambda_{{\rm cr}}$, 
below which the parent cloud is stable even in the linear regime.  

The above conjecture is supported in the planar case  
by the fact that the gravitational potential energy per unit 
length per cell increases in absolute value with the amplitude 
$A$ (S-B 1967).  In the cylinder, the same trend is observed 
in $W$ vs.\ $\rho_c/\rho_s$ (Fig.\ \ref{fig-virial}).  Unfortunately, 
the lack of a 2D analytic solution---and therefore an amplitude---in 
the cylindrical case precludes the derivation of an algebraic relation 
between the intercore separation and $\lambda_{z,{\rm cr}}$.  However, 
a close inspection of the numerical results suggests that it is again 
$\lambda_{z,{\rm cr}}$ that is of primary importance in filament 
fragmentation.  

Recall from the results of \S \ref{sec-numer2} that we were able to 
find 2D equilibria only down to a minimum $Z' \simeq 2.1$, where 
$\rho_c/\rho_s = 1.2$. Below this value, the method converged to the 
S-O solution. 
Now, if the minimum length scale for fragmentation were set by 
the critical wavelength as in the planar case, then we would expect 
to find 2D solutions down to a minimum value of $Z' = Z/l_0 = 
\lambda_{z,{\rm cr}}/(2l_0) \simeq 1.97$ (\S \ref{sec-equi}), with a 
corresponding $\rho_c/\rho_s \simeq 1$. Our minimum converged value 
of $Z' = 2.1$ is consistent with this expectation, given the intrinsic 
inaccuracy of the numerical method.  Moreover, the fact that equilibria 
with $\rho_c/\rho_s > 1.2$ {\it were} found for $2.1 \leq Z' \leq 3.91 
= \lambda_{z,{\rm MGR}}/(2l_0)$ certainly argues {\it against} any possible 
significance of $\lambda_{z,{\rm MGR}}$ in the final equilibria.

Should a subsequent stability analysis of these solutions reveal that 
the mass peak in Fig.\ \ref{fig-plseq} indeed signifies a stability 
transition, then an interesting, although highly approximate, criterion 
for whether a given filament will form stars can be derived.  The argument 
proceeds as follows.  The mass peak of Fig.\ \ref{fig-plseq} corresponds 
to $\td{Z}_\pk = 0.94$, whence $Z_\pk = 0.94~ l_s = 
0.36~(\sigma/a_{10})$ pc.  Thus, $2 Z_\pk = 0.72~(\sigma/
a_{10})$ pc represents the scale of a single whole 
fragment formed out of the parent cloud.  But what is the affected 
region of the original filament?  It is likely to be somewhat larger 
than $2 Z_\pk$, by the argument summarized in the paragraph 
above.  Roughly speaking, the affected region should be larger by a factor 
$\sim \lambda_{z,{\rm MGR}}/
\lambda_{z,{\rm cr}} \simeq 2$, since $\lambda_{z,{\rm cr}}$ gives the 
scale at which the filament first becomes unstable to fragmentation 
(\S \ref{sec-equi}).  
Thus, assuming that the original filament was not radially collapsing 
($\mu < \mu_\mx$), we estimate that any filament longer than $\sim 2 
\cdot 2Z_\pk \simeq 1.44~(\sigma/a_{10})$ pc may be expected 
to form stars via fragmentation and subsequent dynamical collapse of 
the prolate fragments.  Shorter filaments, which could only harbor 
cores with $\rho_c/\rho_s < 10$, would require another agent--- e.g., 
an increase in external pressure, ambipolar diffusion, etc.---to 
initiate collapse. 

\section{CONCLUSION}
We have presented a new family of 2D, numerical solutions of the 
isothermal Lane-Emden equation in cylindrical symmetry.  The equilibria 
have an embedded, periodic structure, and in this sense are direct 
counterparts to the 2D solution of Schmid-Burgk (1967) in planar 
symmetry.  Together, the existence of the two equilibria suggest a 
fragmentation scheme which seems consistent with the observed 
hierarchical structure in several well known star-forming 
regions.  Moreover, our results constitute a remarkably 
simple yet robust explanation for the origin and maintenance of 
the prolate, gaseous cores that represent the lower rung of this 
hierarchy.  Although we have ignored many 
physical effects that should be included in subsequent studies 
(particularly nonthermal motions and magnetic fields), we expect 
that the shift in emphasis from isolated to embedded structures will 
prove fruitful in future theoretical work on the origin and 
evolution of dense cores.  
\\\\
\noi
It is a pleasure to thank Steve Stahler for discussions that prompted 
a search for these solutions.  I am grateful to Richard Larson and Dean 
McLaughlin for useful comments on the manuscript, and to an anonymous 
referee for suggestions which helped to clarify certain properties of 
the equilibria.

\begin{appendix}
\section{RELATION BETWEEN THE SOLUTIONS OF S-B AND S-O}

It is worthwhile noting an intriguing correspondence between the 
solutions of S-B and S-O.  Let us examine the structure of one of 
the cylindrical fragments of \S \ref{sec-plane} in its own right.  
That is, consider the limit $A \goto 1$, so that 
$\rho_{0,s} \propto \ell_0^{-1} \goto 0$, giving an isolated 
cylinder (see equation [\ref{eq-maxmin}]).  Let $\ell_0^{-1} = 
2\pi/\lambda_{x,{\rm cr}} \equiv k_{x,{\rm cr}}$.  Then in the limit 
$k_{x,{\rm cr}} \goto 0$ the solution (\ref{eq-sbsn}) becomes 
\beq
\rho (x,y) = \rho_c \left[ 1 - \frac{A x^2 + y^2}{1 - A} 
k_{x,{\rm cr}}^2 + O(k_{x,{\rm cr}}^4)\right].
\label{eq-lim1} 
\eeq
In terms of $k_{x,{\rm cr}}$, equation (\ref{eq-xamp}) reads
\bed
k_{x,{\rm cr}}^2 \ell_c^2 = \frac{1-A}{1+A}, 
\eed
the substitution of which into equation (\ref{eq-lim1}) gives
\beqa
\rho (x,y) & = & \rho_c \left[ 1 - \frac{A x^2 + y^2}
{\ell_c^2 ~(1 + A)} + O(k_{x,{\rm cr}}^4)\right] \non \\
& = & \rho_c \left[ 1 - \frac{r^2}{2~\ell_c^2} + 
O(k_{x,{\rm cr}}^4)\right], 
\label{eq-lim2} 
\eeqa
in the limit as $A \goto 1$, 
where we have noted that $r^2 = x^2 + y^2$, appropriate to the 
geometry of the isolated cylindrical fragment. 

The S-O solution, on the other hand, reduces to the following in the 
limit of small $r$ (equation [\ref{eq-sosoln}]):
\beq
\rho (r,z) = \rho_{0,f} \left[ 1 - \frac{2~r^2}{l_0^2} + 
O(r^4) \right]. 
\label{eq-lim3} 
\eeq
Equations (\ref{eq-lim2}) and (\ref{eq-lim3}) are identical, at  
order $r^2$, provided that: (i) $l_0^2 = 4 \ell_c^2$; and (ii) 
$\rho_{0,f} = \rho_c$.  Recalling the definitions 
of $\ell_c$ and $l_0$ given in \S \ref{sec-plane} and \ref{sec-fil}, 
if the latter is true, then so is the former.  But equality (ii) does  
indeed hold, since the planar fragment has a topology indistinguishable  
from that of a cylinder in the prescribed limit.  Thus, the S-O solution 
appears as the unique $A \goto 1$ limit of the S-B solution.  
\end{appendix}

\newpage
\bfig
\plotone{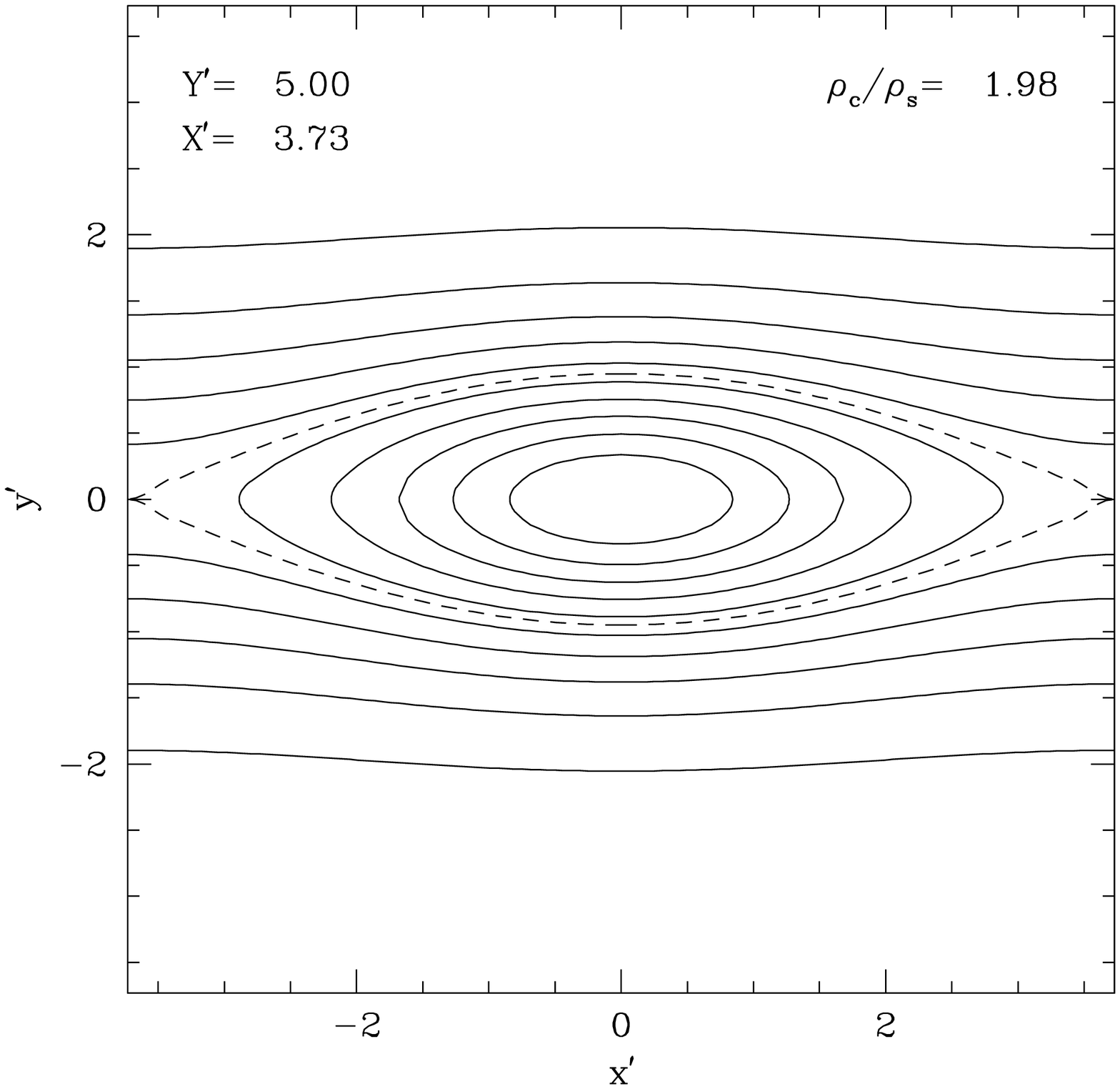}
\caption{The closest numerical equivalent to S-B's $A = 0.17$ solution. 
Solid curves are equidensity contours, linearly spaced in $\rho$ from 
$\rho (0,0)/\rho_s = 1.98$ down to $\rho/\rho_s = 0.18$.  The dashed 
curve indicates the surface $\rho = \rho_s$.  Recall that $x' \equiv 
x/\ell_c, \ldots$ etc. 
\label{fig-sbnum}} 
\efig

\bfig
\plotone{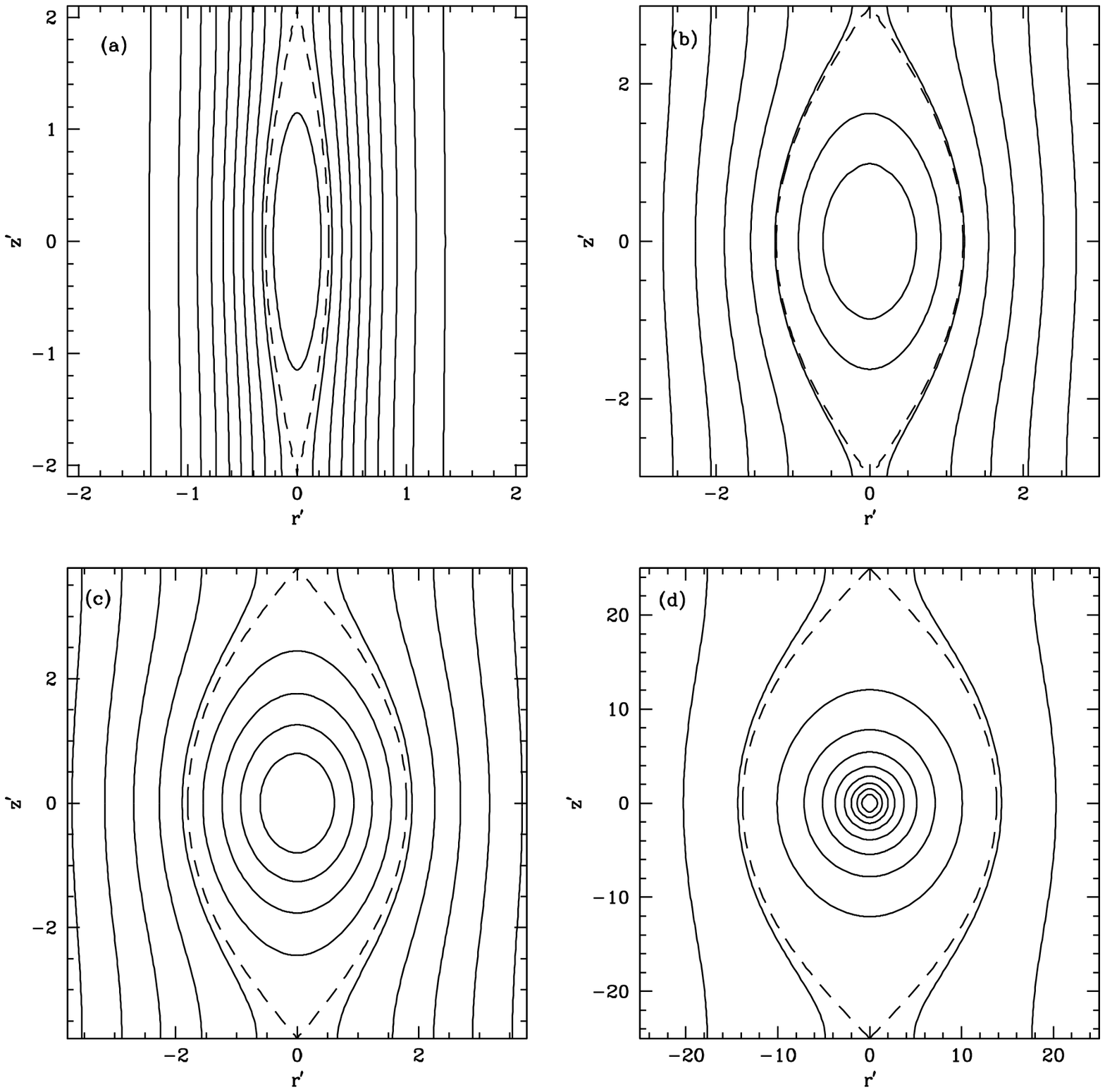}
\caption{Four embedded cores in the isothermal filament.  Each is 
characterized by its value of $(Z',R')$: (a) (2.1, 2.1); (b) (3.0, 5.0); 
(c) (3.79,5.0); (d) (25.0,28.0). 
The corresponding density contrasts, $\rho_c/\rho_s$, are 1.17, 4.70, 
10.49, and 1076, respectively.  Contours are spaced linearly in (a), 
logarithmically in (b), (c), and (d).  Recall that $r' \equiv 
r/l_c, \ldots$ etc. 
\label{fig-plegs}} 
\efig

\bfig
\plotone{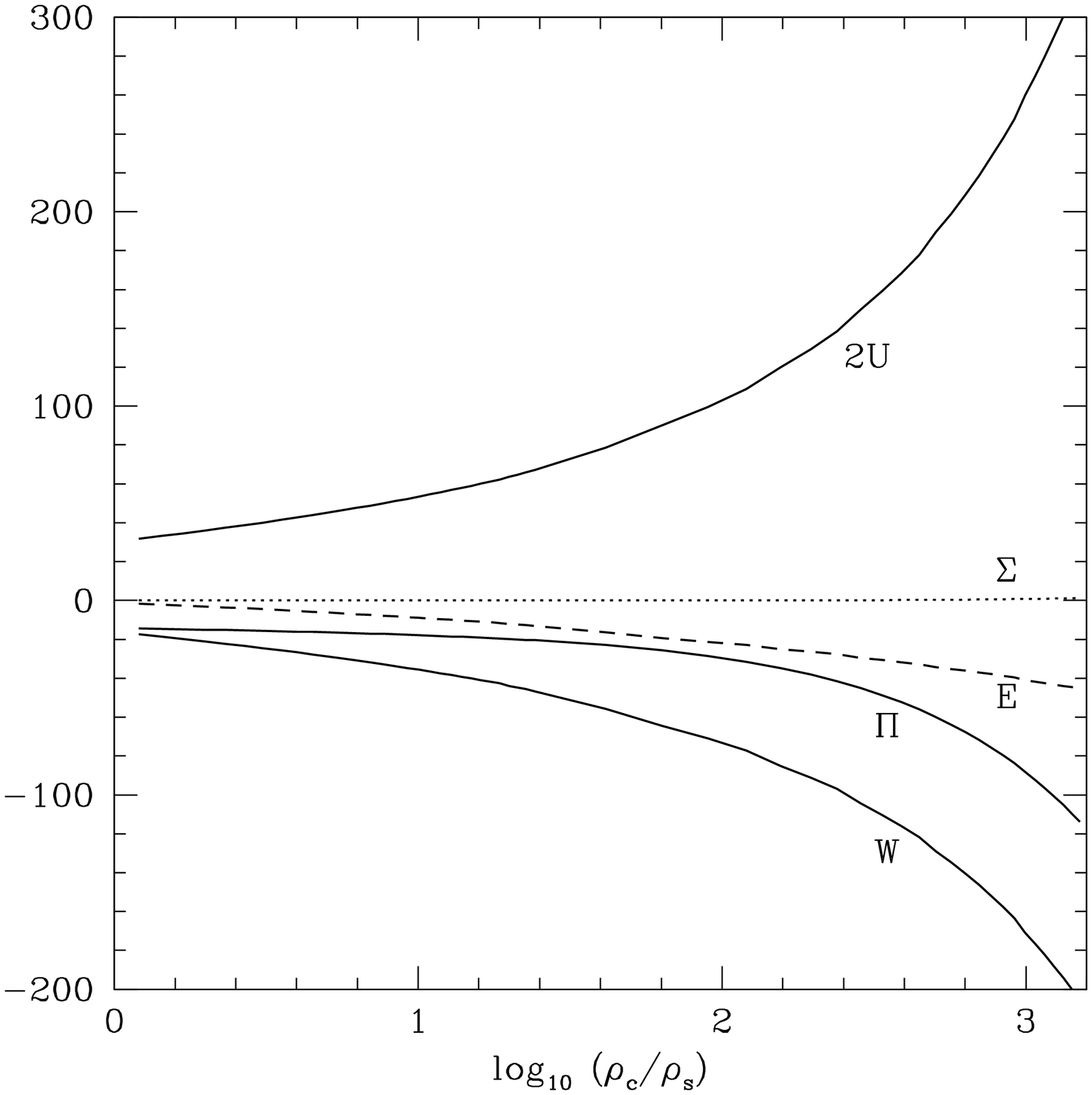}
\caption{
Solid curves represent individual terms of the virial theorem for an 
entire cell, plotted 
against core density contrast (center to tidal lobe).  The dotted curve 
is the sum of these terms, while the dashed curve is the total energy.  
See \S \ref{sec-virial} for details. 
\label{fig-virial}} 
\efig

\bfig
\plotone{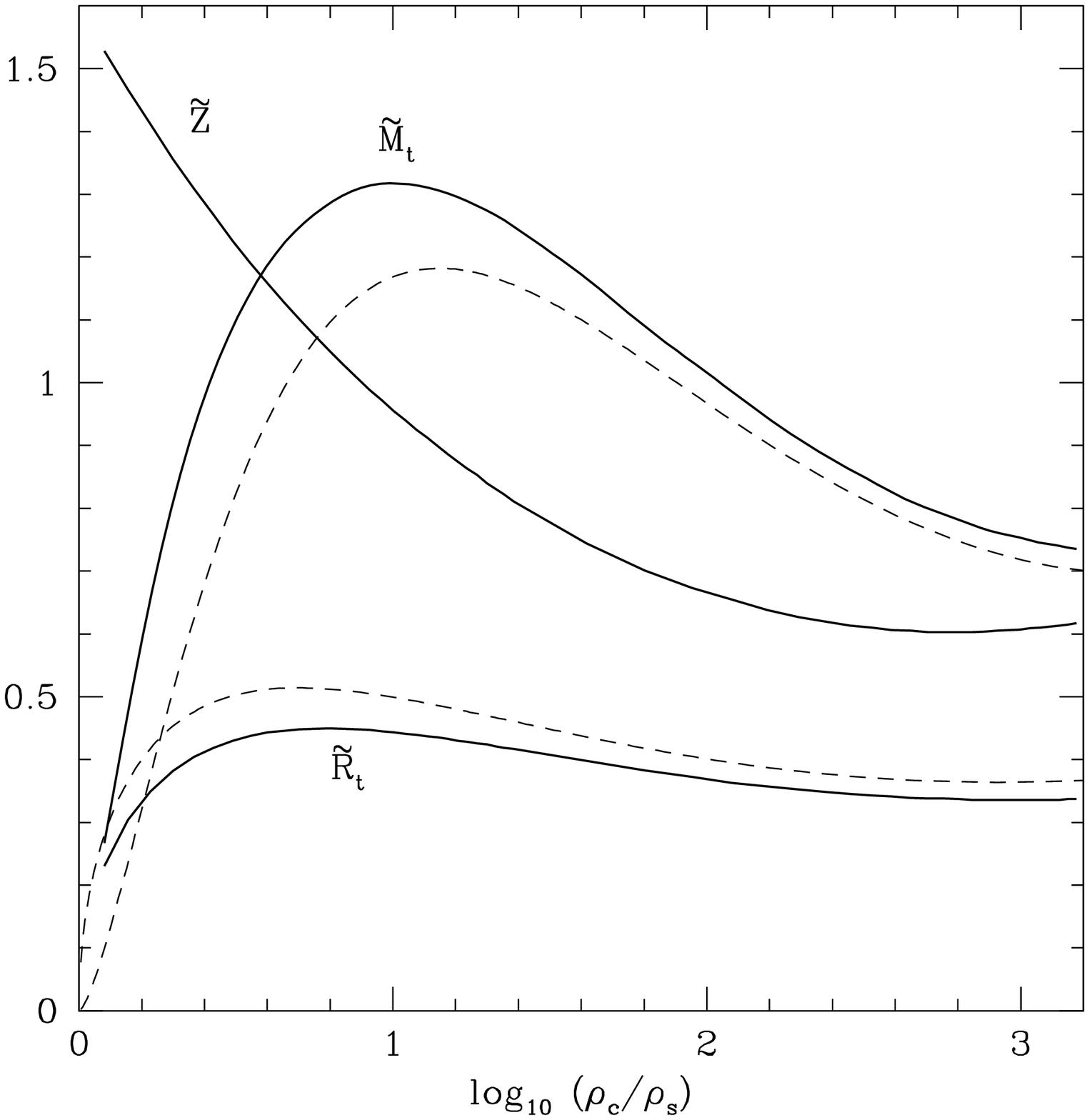}
\caption{Solid curves show the entire sequence of models that 
were calculated, represented as $\td{Z}, \td{R}_t$, and 
$\td{M}_t$ vs.\ central concentration, $\rho_c/\rho_s$.  
The overall behavior is similar to that of the B-E sequence, 
shown by dashed lines (dimensionless mass, top; dimensionless 
radius, bottom).  See text for details. 
\label{fig-plseq}} 
\efig

\begin{figure}[t]
\begin{tabular}{c}
\parbox{7.0cm}{
\epsfxsize=9.0cm
\leavevmode
\hspace*{3.5cm}
\epsffile{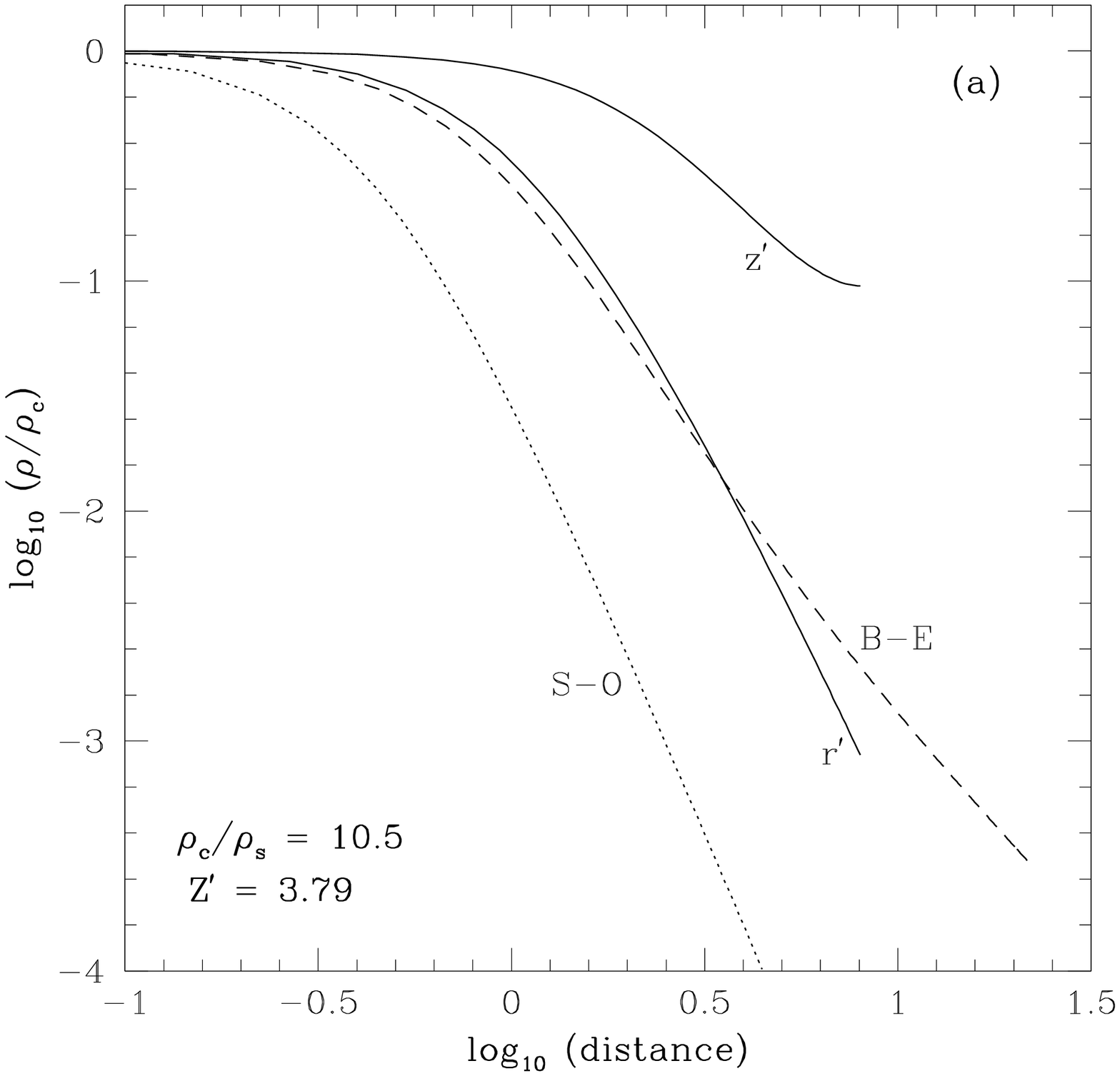}
\vspace*{0.3cm}
}\\
\parbox{7.0cm}{
\epsfxsize=9.0cm
\leavevmode
\hspace*{3.5cm}
\epsffile{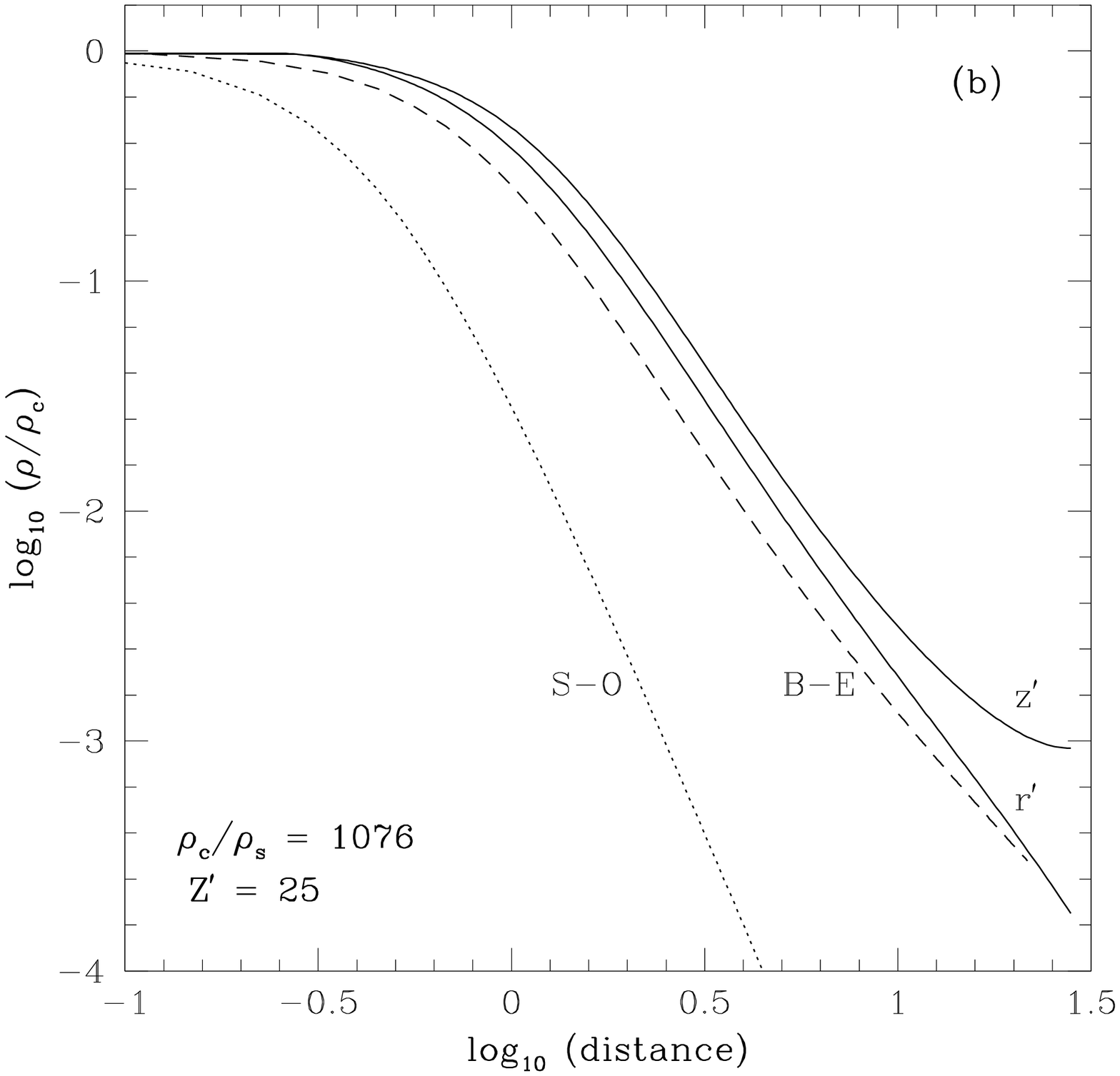}
\vspace*{1cm}
}\\
\end{tabular}
\caption{(a) Density profiles along the polar ($z'$) and equatorial ($r'$) 
axes of the equilibrium depicted in (c) of Fig.\ \ref{fig-plegs}.  
Also shown are the radial profiles of a S-O cylinder (dotted 
curve) and a B-E sphere (dashed curve). (b) Same as (a), for the 
equilibrium depicted in Fig.\ \ref{fig-plegs} (d).
\label{fig-plrho}} 
\efig

\bfig
\plotone{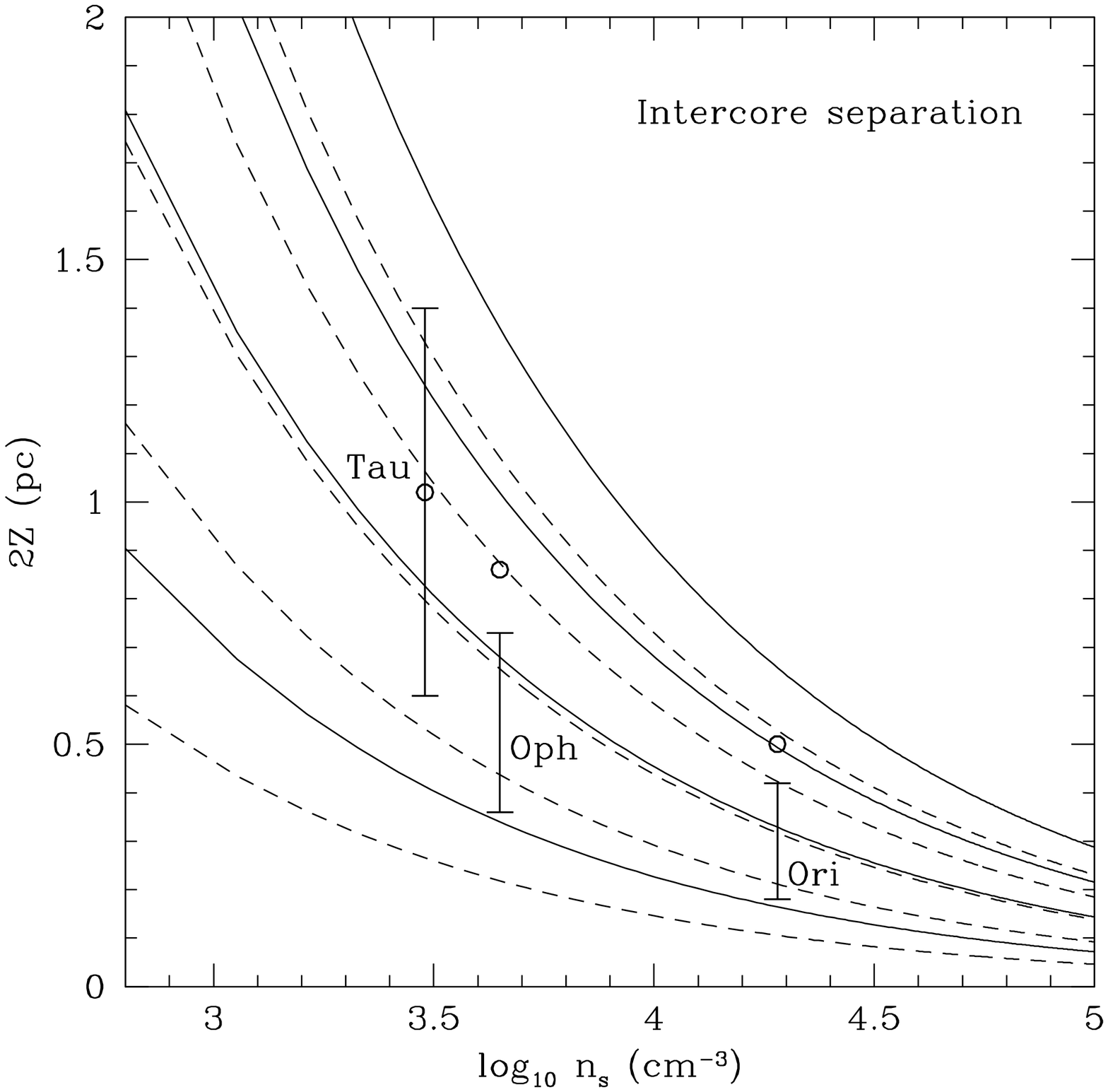}
\caption{Intercore separation, $2Z$, as a function of $n_s$.  
Solid curves are 
$2Z_\pk$; from bottom to top, $\sigma/a_{10} =$ 1, 2, 3 and 4, 
respectively.  Dashed curves are $2Z_\mn$; from bottom to top, 
$\sigma/a_{10} =$ 1, 2, 3, 4 and 5, respectively. 
Vertical segments indicate approximate ranges of $2Z$ 
observed in three regions: from left to right, Taurus, Ophiuchus, and 
Orion.  Open circles are predicted values of $2Z$, assuming no 
inclination of the filament to the line of sight.  See Table 1 and 
text for details.  
\label{fig-plic}} 
\efig

\newpage
\scriptsize
\begin{table}[htb]{\sc Table 1: Mean Properties of Selected Filaments 
in Taurus, Ophiuchus, and Orion.}
\begin{center}
\begin{tabular}{lcccccccc}
\tableline\tableline
\noalign{\vskip0.05truein}
~~~~Region,~~~~ & ~~$\mu_\obs$~~ & 
~~log$_{10}~n_{s,\obs}$ ~~ & ~~$\sigma_{\obs}$~~ & 
~~$2Z_\obs$~~ & 
~~$2Z_{\rm pred}$~~ & ~~$\sigma_\mx$~~ & Refs. \\
~~~~Cloud~~~~ & ~~($M_{\sun}/$pc)~~ & ~~(cm$^{-3}$)~~ & 
~~(km s$^{-1}$)~~ & ~~(pc)~~& ~~(pc)~~& 
& ~~(km s$^{-1}$) \\
\noalign{\vskip0.05truein}
\tableline
\noalign{\vskip0.05truein}
Taurus, & ~ & ~ & ~ & ~ & ~ & ~ & ~ \\
L1495 & 100 & 3.48 & 0.47 & 0.6---1.4 & 1.02 &
0.47 & 1,2; 2; 3; 4 \\ 
~ & ~ & ~ & ~ & ~ & ~ & ~ & ~ & ~ \\
Ophiuchus, & ~ & ~ & ~ & ~ & ~ & ~ & ~ \\
L204 & 110 & 3.65 & 0.48 & 0.36---0.73 & 0.86 &
0.49 & 5; 5; 5; 6 \\ 
~ & ~ & ~ & ~ & ~ & ~ & ~ & ~ & ~ \\
Orion, & ~ & ~ & ~ & ~ & ~ & ~ & ~ \\
L1641 & 385 & 4.28 & 0.58 & 0.18---0.42 & 0.50 &
0.92 & 7; 8; 8; 9 \\ 
\noalign{\vskip0.05truein}
\tableline
\noalign{\vskip0.05truein}
\end{tabular}
\end{center}
\tablecomments{
The filaments in each region were chosen on the basis of their large 
apparent axial ratios (length, $L$/width $\gae 20$) as observed in 
$^{13}$CO maps, from which both $M$ and $L$, and thus $\mu$, were 
deduced.  The intercore separations, $2Z_\obs$, are nearest core 
distances derived from C$^{18}$O maps in Taurus and Ophiuchus and 
NH$_3$ maps in Orion.  The mean intercore density, $n_{s,\obs}$, is 
the density corresponding to the mean $2Z_\obs$ in each 
region.  This requires results from the appropriate (scale-dependent) 
molecular tracer: $^{13}$CO in Taurus and Ophiuchus, and C$^{18}$O in 
Orion.  The velocity dispersion, $\sigma_{\obs}$, is chosen in the 
same manner.  References to the literature are ordered as follows.  
The first reference in each row is the source of $\mu_\obs$, followed 
by references to the sources of $n_{s,\obs}, ~\sigma_{\obs}$, and 
$2Z_\obs$, each separated by a semicolon.}
\tablerefs{(1) Gaida et al 1984; 
(2) Mizuno et al 1995; (3) Nercessian et al 1988; (4) Onishi et al 1996; 
(5) Nozawa et al 1991; (6) Tachihara et al 2000; (7) Bally et al 1987; 
(8) Dutrey et al 1993; (9) Cesaroni \& Wilson 1994.}
\label{tab-table}
\end{table}

\end{document}